\documentclass[fleqn,usenatbib]{mnras}

\usepackage{newtxtext,newtxmath}

\usepackage[T1]{fontenc}

\DeclareRobustCommand{\VAN}[3]{#2}
\let\VANthebibliography\thebibliography
\def\thebibliography{\DeclareRobustCommand{\VAN}[3]{##3}\VANthebibliography}

\usepackage{graphicx}	
\usepackage{amsmath}	
\usepackage{xcolor}

\newcommand{\isvec}[1]{\bf{#1}}

\title[Shock Variability of low-$q$ SMBHBs]{Shock-Driven Periodic Variability in a Low-Mass-Ratio Supermassive Black Hole Binary}

\author[K. Whitley et al.]{
K. Whitley,$^{1}$\thanks{E-mail: kwhitle@umich.edu}
A. Kuznetsova,$^{2}$
K. G{\"u}ltekin,$^{1}$
and M. Ruszkowski$^{1}$
\\
$^{1}$Department of Astronomy, University of Michigan, 1085 South University Avenue, Ann Arbor, MI 48109-1107, USA\\
$^{2}$Department of Astrophysics, American Museum of Natural History, 200 Central Park West, New York, NY 10024-5102, USA\\
}

\date{Accepted XXX. Received YYY; in original form ZZZ}

\pubyear{2023}

\begin{document}
\label{firstpage}
\pagerange{\pageref{firstpage}--\pageref{lastpage}}
\maketitle

\begin{abstract}
    We investigate the time-varying electromagnetic emission of a low-mass-ratio supermassive black hole binary (SMBHB) embedded in a circumprimary disk, with a particular interest in variability of shocks driven by the binary.
    We perform a 2D, locally isothermal hydrodynamics simulation of a SMBHB with mass ratio $q=0.01$ and separation $a=100\;R_g$, using a physically self-consistent steady disk model. We estimate the electromagnetic variability from the system by monitoring accretion onto the secondary and using an artificial viscosity scheme to capture shocks and monitor the energy dissipated.
    The SMBHB produces a wide, eccentric gap in the disk, previously only observed for larger mass ratios, which we attribute to our disk model being much thinner ($H/R\approx0.01$ near the secondary) than is typical of previous works. The eccentric gap drives periodic accretion onto the secondary SMBH on a timescale matching the orbital period of the binary, $t_{\rm{bin}}\approx0.1\;\rm{yr}$, implying that the variable accretion regime of the SMBHB parameter space extends to lower mass ratios than previously established. Shocks driven by the binary are periodic, with a period matching the orbital period, and the shocks are correlated with the accretion rate, with peaks in the shock luminosity lagging peaks in the accretion rate by $0.43\;t_{\rm{bin}}$. We propose that the correlation of these quantities represents a useful identifier of SMBHB candidates, via observations of correlated variability in X-ray and UV monitoring of AGN, rather than single-waveband periodicity alone.
\end{abstract}




\section{Introduction}

Since most galaxies are host to a supermassive black hole (SMBH) \citep{kh13} and galaxy mergers are a regular occurrence \citep{lotz11}, it is expected that there should exist a population of post-merger galaxies containing a supermassive black hole binary (SMBHB) \citep{begelman1980}. These binaries are a necessary precursor to SMBH mergers, and both binaries and mergers are sources of gravitational waves (GWs) which may be detected by Pulsar Timing Array (PTA) collaborations and the upcoming Laser Interferometer Space Antenna (LISA) mission \citep{lisa17}, respectively. While there is no detection yet for a continuous-wave GW signal of an individual binary \citep{nanograv23}, the NANOGrav collaboration has reported evidence for a stochastic common-process signal, which is expected for a gravitational wave background (GWB) generated by a population of SMBHBs \citep{middleton16}, though the current evidence remains insufficient to claim detection of a GWB \citep{nanograv20}.

These hints and future GW detections can best be used for astrophysical inference when combined with electromagnetic (EM) observations, necessitating the development of methods for identifying and studying candidate SMBHBs with EM data alone. At wider separations, candidate multiple-SMBH systems can be identified through spatially resolving them \citep[down to $\sim$1 kpc;][]{2003ApJ...582L..15K, comerford12, blecha13, foord19}, through HI absorption lines with different Doppler shifts implying the presence of a second relativistic jet (down to $\sim$few pc) \citep{rod09,2016MNRAS.459..820T,2017ApJ...843...14B}, or through identifying spectrally distinct broad line emission regions (down to $\sim$0.1 pc) \citep{eracleous12, runnoe17}. However, at the very close ($\ll1$ pc) separations most relevant to multimessenger studies with GWs, we rely on searches for active galactic nuclei (AGN) with periodically-modulated lightcurves \citep[among others]{graham15,liu19,liu20,chen20}. Unfortunately, even single-SMBH AGN are intrinsically variable sources, such that distinguishing periodicity from stochastic variability requires observing many cycles of the hypothetical period \citep{vaughan16}. It is important, then, to understand the properties of binary-induced EM periodicity which distinguish it from other sources of AGN variability.

A great deal of numerical simulations have been performed to understand the nature of SMBHB variability and evolution, investigating trends with binary mass ratio, separation, and eccentricity, as well as with disk temperature and viscosity \citep{mm08,dorazio13,farris14,farris15,dorazio16,tang17,tang18,moody19,tiede20,duffel20,ws22,dorazio21,derd21}.

The vast majority of the existing literature has used a locally isothermal equation of state, with accretion rate onto the binary components being used as the proxy for EM variability. However, \citet{tang18} have shown in a non-isothermal simulation that shock-heating of the cavity wall produces significant periodicity at high energies. This shock-induced periodicity can still be monitored in locally isothermal simulations via shock capturing, as binary-induced shocks are still present and dissipating energy, offering a second proxy for EM variability to analyze alongside accretion rates. 

A particularly interesting case to examine is low-$q$ SMBHBs. The majority of galaxy mergers are unequal-mass, and thus are expected to produce unequal-mass SMBHBs, with the caveat that smaller BHs are likely to form binaries less efficiently. Additionally, there are other proposed formation channels for low-$q$ SMBHBs, such as the presence of low-mass ($10^2$--$10^4$ M$_\odot$) SMBH seeds \citep{bellovary2011}, in-situ growth of stellar mass BHs to intermediate mass black hole (IMBH) masses in the disk \citep{insitu12}, and disruption of IMBH-hosting globular clusters \citep{2016MNRAS.458.2596F,2018ApJ...856...92F,2018MNRAS.477.4423A,2018ApJ...867..119F,2019ApJ...878...17R}. However, it has been found previously that accretion onto the binary components becomes steady for $q<0.05$ \citep{dorazio16}. It is of interest, then, to determine whether periodic shocks also vanish for low-$q$ SMBHBs, or whether the parameter space for EM-variable SMBHBs extends to lower mass ratios than previously established.

In this work, we perform a 2D hydrodynamic simulation of a $q=0.01$ SMBHB and examine its implied variable electromagnetic properties via monitoring of both accretion and shocks. Shock capturing is performed using an artificial viscosity to both identify shocks and calculate the energy dissipated by them \citep{vnr}. Further, we relax the common assumption of constant disk aspect ratio in favor of a physically self-consistent disk model in order to better reproduce the real environments of these systems.

This paper is organized as follows. In Section \ref{sec:methods}, we describe the simulation setup, including the models used for the disk, gravity, accretion, and shock capturing. Section \ref{sec:results} presents our analysis of the disk dynamics and morphological evolution and our monitored accretion rate and shock outputs. We discuss the implications of these results in Section \ref{sec:discussion}, in particular the dynamical and morphological differences from past works at this mass ratio (\S\ref{sec:disc:dynamics}) and EM observables (\S\ref{sec:disc:variability}). Section $\ref{sec:conclusion}$ summarizes our findings. Appendix \ref{appendix:disk} provides a derivation for the disk model used in this work.

\section{Methods}\label{sec:methods}

In this section, we describe the setup of the numerical simulation performed for this work. We review the initial conditions used for the system, the physics included in the simulation, and methodology for monitoring accretion and shock dissipation, which will serve as proxies for electromagnetic variability in our analysis.

\subsection{FARGO3D}

The simulations run for this work were performed on a two-dimensional, Eulerian, cylindrical grid using the \emph{FARGO3D} code \citep{fargo3d}. \emph{FARGO3D} solves the hydrodynamics equations 

\begin{equation}
    \frac{\partial \Sigma}{\partial t} + \nabla \cdot \left(\Sigma \isvec{v}\right) = 0,
\end{equation}
\begin{equation}
    \Sigma \left( \frac{\partial \isvec{v}}{\partial t} + \isvec{v} \cdot \nabla\isvec{v}\right) = -\nabla P - \Sigma \nabla \Phi + \nabla\cdot\Pi, 
\end{equation}
where $\Sigma$ is the surface density, $\isvec{v}$ is the fluid velocity, $P=\Sigma c_s^2$ is the isothermal pressure with $c_s$ the isothermal sound speed, $\Phi$ is the gravitational potential, and $\Pi$ is the viscous stress tensor. We use \emph{FARGO3D}'s built-in $\alpha$-prescription to set the viscosity as $\nu = \alpha c_s H$ \citep{ss73}, where $H$ is the local scale height of the disk.

For the transport step in the azimuthal direction, \emph{FARGO3D} employs the FARGO orbital advection algorithm \citep{fargo}, which greatly increases the solution accuracy and allowable simulation timestep for systems dominated by rotation, such as accretion disks.

\subsection{Disk Setup} \label{methods:setup}

Our simulated domain is that of a disk centered on a primary SMBH with mass $M_1=10^8\ M_\odot$ and extends from $R_\mathrm{min}=10\;R_g$ to $R_\mathrm{max}=1000\;R_g$, where $R_g=GM_1/c^2$ is the gravitational radius of the primary SMBH. The secondary SMBH has mass $M_2=10^6\  M_\odot$ and is placed at a separation $a=100\;R_g$ from the primary, resulting in a binary orbital timescale $t_{\rm{bin}}\approx0.1\;\rm{yr}$. We use the torque-free boundary conditions described by \citet{dempsey2020} for our inner boundary, including the use of the \citet{deval06} wave-killing prescription, but allow outflow at the outer boundary with no wave-killing.

\begin{figure*}
    \center
    \includegraphics[trim= 0pt 24pt 0pt 0pt, clip,width=1.0\textwidth]{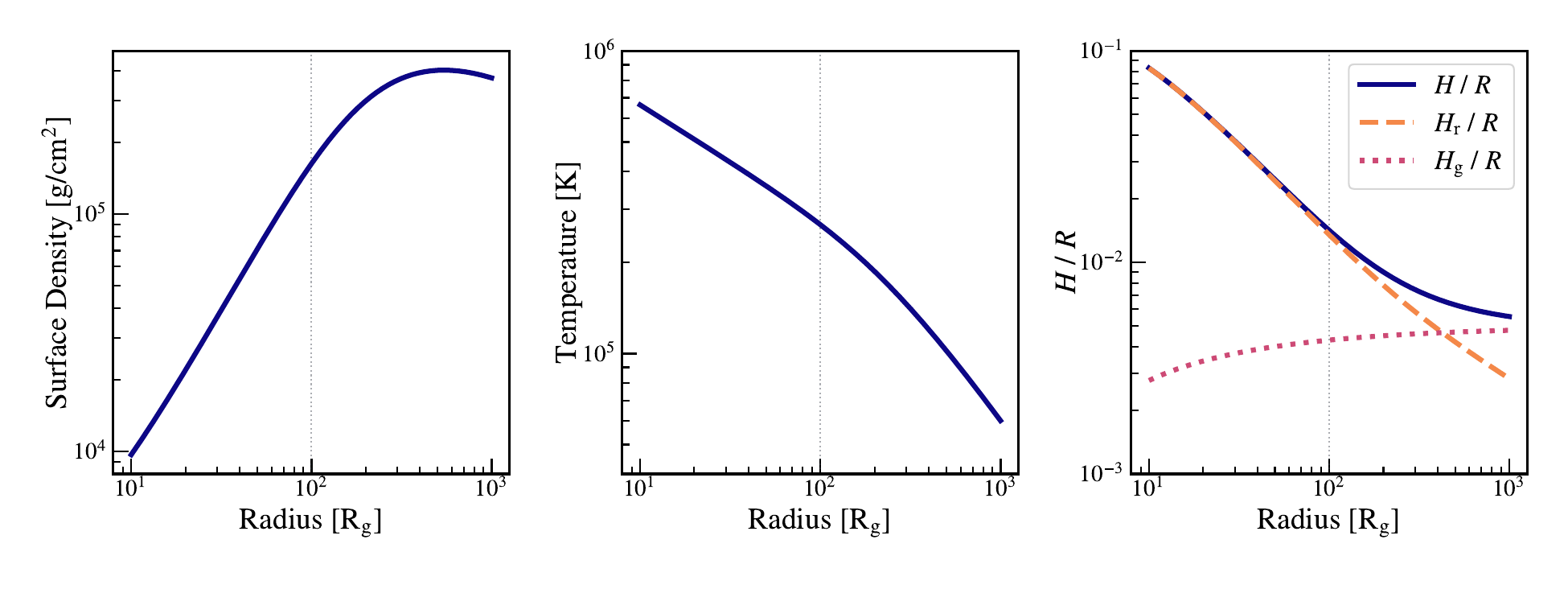}
    \caption{The initial conditions used for the simulation, calculated from the disk model described in Section \ref{methods:setup}. The vertical dotted line on each panel shows the position of the secondary SMBH. For $H/R$ (\emph{solid purple}), $H_r/R$ (\emph{dashed orange}) and $H_g/R$ (\emph{dotted magenta}), where $H_r^2\equiv P_r/(\Sigma \Omega_K^2)$ and $H_g^2\equiv P_g/(\Sigma \Omega_K^2)$, are also plotted to illustrate the relative contributions of radiation and gas pressure, respectively, throughout the disk.}
    \label{fig:ics}
\end{figure*}

The initial conditions of the disk are derived from the steady disk equations \citep{fkr},

\begin{equation}\label{eq:steadydisk}
    \left.
    \begin{aligned}
        &\rho = \Sigma / H;\\
        &H = c_s R^{3/2}/\left(GM\right)^{1/2};\\
        &c_s^2 = P/\rho;\\
        &P = \frac{\rho k_B T_c}{\mu m_p} + \frac{4\sigma}{3c} T_c^4;\\
        &\frac{4\sigma T_c^4}{3\tau} = \frac{3GM\dot{M}}{8\pi R^3}\left[1 - \left(\frac{2R_g}{R}\right)^{1/2}\right];\\
        &\tau = \Sigma \kappa_{\rm R}\left(\rho,T_c\right)=\tau\left(\Sigma,\rho,T_c\right);\\
        & \nu\Sigma = \frac{\dot{M}}{3\pi} \left[1-\left(\frac{2R_g}{R}\right)^{1/2}\right];\\
        &\nu = \nu\left(\rho,T_c,\Sigma,\alpha,...\right),
    \end{aligned}
    \right\},
\end{equation}
where $\rho$ is the gas density, $\Sigma$ is the surface density, $H$ is the scale height, $c_s$ is the sound speed, $R$ is the radial distance from the central object, $M$ is the mass of the central object, $\dot{M}$ is the accretion rate through the disk, $P$ is the total pressure, $T_c$ is the temperature at the disk midplane, $\nu$ is the viscosity, $\tau$ is the optical depth, and $\kappa_{\rm R}$ is the Rosseland mean opacity, which throughout this work is assumed to be dominated by electron scattering, $\kappa_{\rm R}\approx\kappa_{\rm e}=0.4\;\mathrm{cm^2/g}$. We choose an $\alpha$-viscosity, $\nu=\alpha c_s H$ \citep{ss73}, and introduce a prescription for the scale height in order to control where the disk transitions from radiation-dominated to gas-dominated and the proportionality of $H$ to $r$ in the gas-dominated region:

\begin{equation}\label{eq:scaleheight}
    H(r) = h(r)  d(r)^{1/2};
\end{equation}
\begin{equation}\label{eq:minheight}
    h(r) \equiv \frac{3}{2} \frac{GM_\odot}{\eta c^2} m\dot{m}\left[1-\left(\frac{2R_g}{R}\right)^{1/2}\right],
\end{equation}
\begin{equation}
    d(r) \equiv \left[\left(\frac{r}{r_c}\right)^{2k} + \frac{1}{2}\left(1+\sqrt{1+4\left(\frac{r}{r_c}\right)^{2k}}\right)\right].
\end{equation}
Here, $m\equiv M/M_\odot$ is the mass of the central object in units of solar masses, $r\equiv R/R_g$ is the radial coordinate in units of gravitational radii ($R_g\equiv GM/c^2$), and $\dot{m}\equiv \dot{M}/\dot{M}_\mathrm{Edd}$ is the accretion rate in units of Eddington, where $\dot{M}_\mathrm{Edd}\equiv 4\pi GMm_p/(\eta \sigma_T c)$ with $m_p$ being the mass of a proton, $\sigma_T$ the Thomson cross-section, and $\eta\equiv L/(\dot{M}c^2)$ the accretion efficiency. The parameters $r_c$ and $k$ set, respectively, the radius at which the disk pressure transitions from radiation to gas-dominated and the powerlaw slope of $H$ in the gas-dominated region (i.e., $H \propto r^{k}$ far from the central object). With these prescriptions for $\nu$ and $H$ in hand, we are able to solve the system in equation~(\ref{eq:steadydisk}) for the surface density and sound speed distributions,

\begin{equation}\label{eq:density}
    \Sigma(r) = \frac{16}{27}\frac{m_p}{\sigma_T} \eta \alpha^{-1} \dot{m}^{-1} r^{\frac{3}{2}} \left(1 - \sqrt{\frac{2}{r}}\right)^{-1} d\left(r\right)^{-1},
\end{equation}

\begin{equation}\label{eq:soundspeed}
    c_s(r) = \frac{3}{2}c\eta^{-1}\dot{m}r^{-\frac{3}{2}}\left(1 - \sqrt{\frac{2}{r}}\right)^2 d\left(r\right)^{\frac{1}{2}}.
\end{equation}

A full derivation of these expressions (equations~(\ref{eq:scaleheight})--(\ref{eq:soundspeed})) can be found in Appendix \ref{appendix:disk}. For this work, we use $\alpha=10^{-2}$, $\dot{m}=10^{-1}$, $\eta=10^{-1}$, $r_c=300$, and $k=1$. Fig.~\ref{fig:ics} shows the surface density, temperature, and scale height distributions calculated from equations~(\ref{eq:density}), (\ref{eq:soundspeed}), and (\ref{eq:scaleheight}) for our initial conditions. Note that throughout this derivation we consider $P=P_{\rm{gas}}+P_{\rm{rad}}$, the sum of the gas and radiation pressures. The resulting $c_s(r)$ from equation~(\ref{eq:soundspeed}) used in our simulation is then an effective isothermal sound speed which includes the contributions from both the gas and radiation pressure.

The simulation is run on a cylindrical grid of $N_r=2560,\;N_\theta=3493$ cells. The radial spacing of the grid is logarithmic, with $\Delta R/R = 0.0018$. This is chosen in order to resolve the 2:1 eccentric corotation resonance (ECR), which is necessary to correctly capture the evolution of eccentricity in the disk \citep{teyss17}.

\subsection{Gravity}
The regions of interest to this work, namely the secondary's minidisk and the gas comprising and surrounding the gap, are non-relativistic, and so we use a purely Newtonian potential for each black hole. We apply a softening length based on the local disk scale height, giving each black hole a potential of the form
\begin{equation}
    \Phi_i = \frac{GM_i}{\sqrt{r_i^2 + \xi^2 H^2}},
\end{equation}
where $M_i$ is the mass of black hole $i$, $r_i$ is the distance between black hole $i$ and the point of interest, $H=c_s/\Omega_K$ is the disk scale height at that point, with $\Omega_K$ the local Keplerian angular velocity, and $\xi$ is a constant parameter controlling the magnitude of the softening. We choose $\xi=0.6$, following from considerations made by \citet{duffell13}. The calculations do not include the self-gravity of the disk, as the disk is much less massive than the secondary SMBH ($M_\mathrm{disk}/M_2 \sim 0.04$). The mass of the secondary is increased from $0$ to $M_2$ over the first 20 orbits of the simulation in order to soften spurious waves generated by the introduction of this asymmetry to the initially symmetric system.

We set the binary on a fixed, circular orbit. The disk is less massive than the secondary SMBH and so torques from the surrounding gas are not expected to significantly alter the orbit over the duration of the simulation (verified in \S\ref{sec:disc:binorb}). Likewise, we do not expect significant orbital evolution due to the emission of gravitational waves. We can demonstrate this by calculating the ratio of the simulation time to the gravitational inspiral time \citep{pm63},

\begin{equation}\label{eq:inspiral}
    \frac{t_{\mathrm{sim}}}{t_{\mathrm{GW}}} = \frac{512 \pi}{5} N_{\mathrm{orb}} \tilde{r}^{-5/2} q (1 + q)^{1/2} (1 - f_\mathrm{sep}^4)^{-1},
\end{equation}
where $N_{\mathrm{orb}}=10^3$ is the number of orbits in the simulation, $\tilde{r}=10^2$ is the initial separation in units of $R_g$, $q=10^{-2}$ is the mass ratio, and $f_\mathrm{sep}$ is the fraction of the initial separation to which the orbit decays. For a small decay to $f_\mathrm{sep}=0.99$ and our binary parameters, this ratio is below unity, indicating that the orbit will decay by $<1\%$ of the initial separation over the course of our simulation.

\subsection{Accretion Model}\label{sec:m:accr}

The purpose of including accretion of gas onto the secondary SMBH is twofold: first, the removal of gas from the minidisk is a natural part of the disk accretion flow, preventing spurious buildup of material at the location of the secondary; second, the mass accretion rate of an $\alpha$-disk is directly linked to its radiative emission, so monitoring this property allows an estimation of the minidisk's luminosity over time. The secondary removes gas from within an accretion radius $r_{\rm{accr}}=0.5r_{\rm{hill}}$, where $r_{\rm{hill}}=a(q/3)^{1/3}$ is the Hill radius of the secondary, on an accretion timescale, $t_{\rm{accr}}$. $t_{\rm{accr}}$ varies as a function of distance $r$ from the secondary SMBH, based on the viscous timescale for our disk model,
\begin{equation}
    t_{\rm{accr}} = \frac{4}{9} \frac{GM_\odot}{c^3} \eta^2 \alpha^{-1}  m \dot{m}^{-2} r^{7/2} \left(1 - \sqrt{\frac{2}{r}}\right)^2 d\left(r\right), 
\end{equation}
where we use the same values as in our initial conditions for all parameters except that $m=10^6$ and $r$ is now measured in gravitational radii of the secondary. A disk-like accretion timescale such as this steeply increases with distance from the central body, causing accretion to be dominated by the inner parts of the minidisk and relatively insensitive to the specific value of $r_{\rm{accr}}$ chosen.

The fraction of mass removed from cells within $r_{\rm{accr}}$ is computed as 
\begin{equation}
    f_{\rm{accr}} = \frac{\Delta t}{t_{\rm{accr}}},
\end{equation}
where $\Delta t$ is the simulation timestep. We monitor mass, momentum, and angular momentum accreted onto the secondary, binned over every $10^{-3}\;t_{\rm{bin}}$.

\subsection{Shock Capturing}\label{sec:m:shocks}
For shock capturing and characterization, we adapt FARGO3D's von Neumann-Richtmyer artificial viscosity \citep{vnr}, using the tensor implementation described in Appendix B of \citet{zeus2d}. This artificial viscosity has the form
\begin{equation}
    \isvec{Q} =
    \begin{cases}
l^2\Delta x^2 \Sigma \nabla\cdot\isvec{v}\left[\nabla\isvec{v}-\frac{1}{3}\left(\nabla\cdot\isvec{v}\right)\isvec{e}\right],& \text{if } \nabla\cdot\isvec{v}<0\\
0,& \text{otherwise},
    \end{cases}
\end{equation}
where $\Delta x$ is the zone size, $\nabla\isvec{v}$ is the symmetrized velocity gradient tensor, $\isvec{e}$ is the unit tensor, and $l$ is a parameter controlling the number of zone widths a shock is spread over. In this work, we use $l=5$. The key properties of the artificial viscosity are that it broadens shocks such that they are resolved on the grid, is sensitive only to compressive flows ($\nabla\cdot\isvec{v}<0$), and is large inside shocks, but small elsewhere.

In this work, we estimate the EM radiation emitted by shocks from the energy dissipated by the shock capturing scheme, $\partial e/\partial t = -\isvec{Q}:\nabla\isvec{v}$. 
This estimate assumes that shock energy is radiated efficiently, such that the thermal energy escapes the disk and does not significantly contribute to heating the gas on long enough timescales to affect the disk evolution. 

For the range of typical shock Mach numbers in the simulation ($\mathcal{M}\gtrsim10$), the thermal timescale due to radiative diffusion $t_{\rm{diff}} = H^2 (3 C_p / 16\sigma)(\rho^2 \kappa_R / T^3)$ \citep{kw1990}, with $C_p$ the specific heat at constant pressure, is less than $t_{\rm{bin}}$ in the region of interest $R < 300\;R_g$, supporting this assumption.

Like the accretion monitoring, we bin the shock emission over every $10^{-3}\;t_{\rm{bin}}$. The Hill sphere of the secondary is excluded from the analysis of this monitoring.

\section{Results}\label{sec:results}

In this section, we present outputs from the simulation as well as our analysis of the dynamics and of the accretion rate and shock dissipation, which stand in as predictors of the electromagnetic variability of the system. When indicated, we restrict our analysis to simulation outputs after the system has reached a quasi-steady state, when $t > 700\ t_{\rm bin}$, the angular momentum deficit (AMD) has saturated, and the $r>a$ shock luminosity and average accretion rate onto the secondary have plateaued.

\subsection{Disk Morphology and Dynamics}\label{sec:results:dynamics}

\begin{figure*}
    \center
    \includegraphics[trim= 2pt 24pt 2pt 2pt, clip,width=0.47\textwidth]{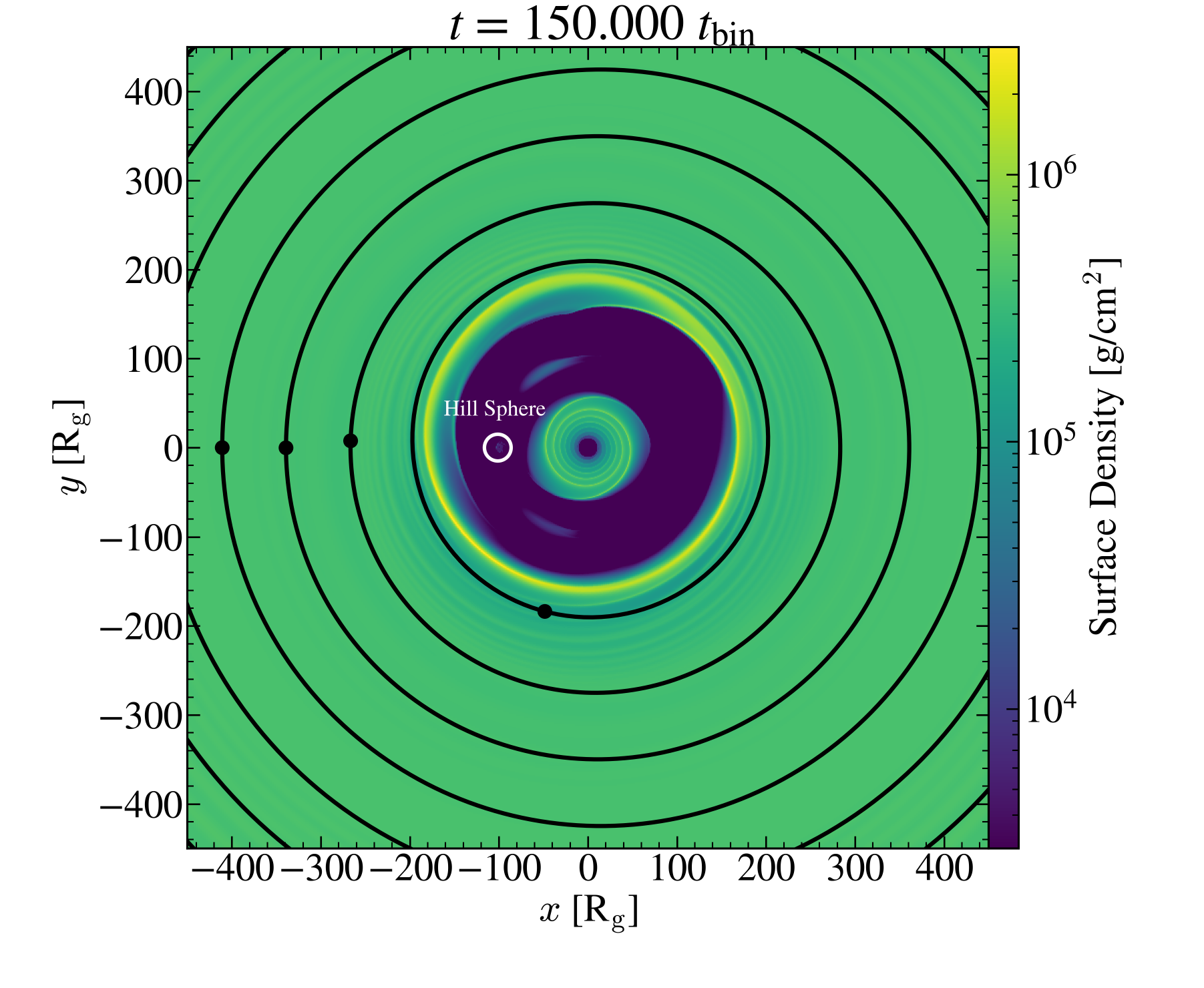}\hspace{-12pt}\includegraphics[trim= 2pt 24pt 2pt 2pt, clip,width=0.47\textwidth]{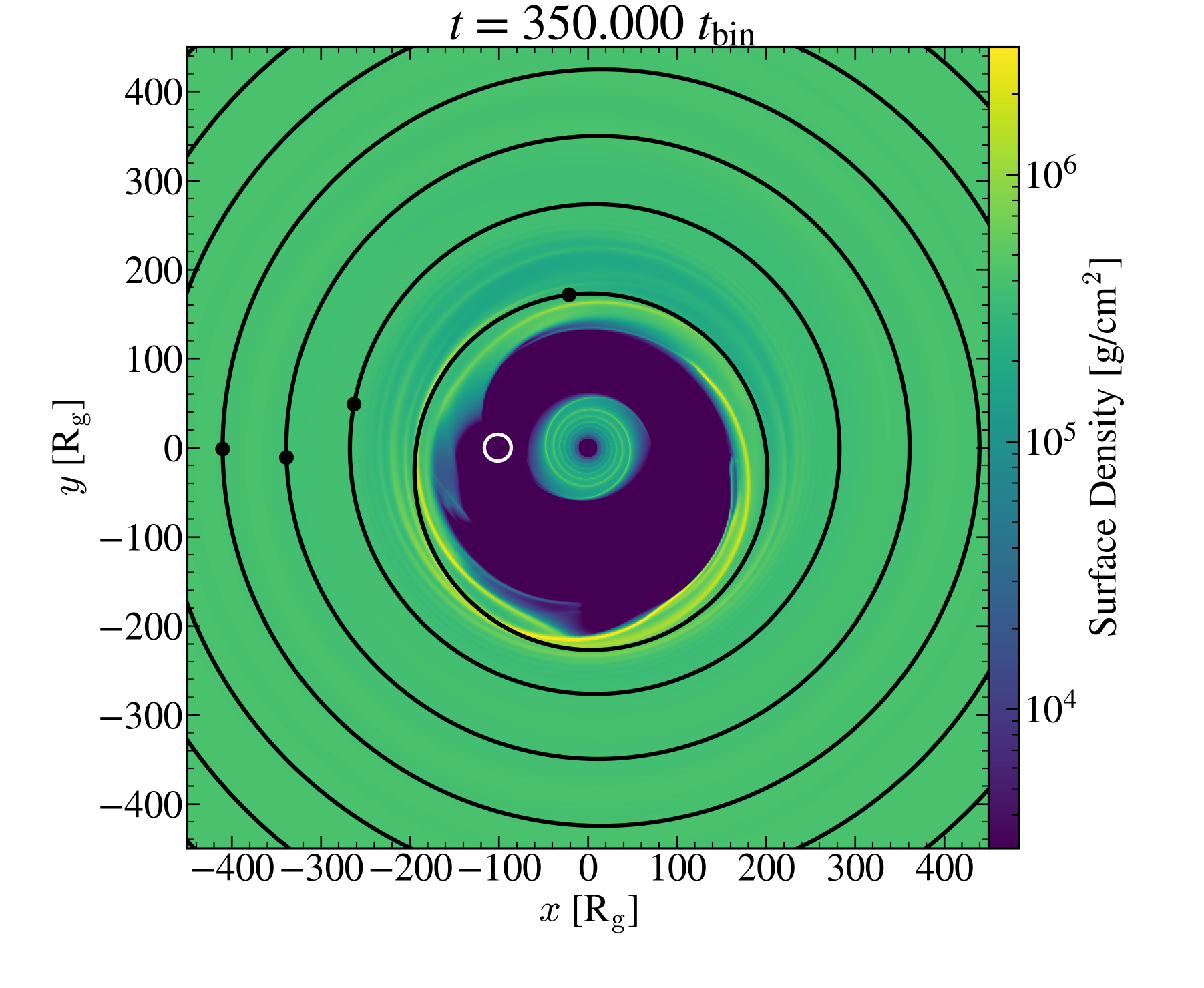}
    \includegraphics[trim= 2pt 24pt 2pt 2pt, clip,width=0.47\textwidth]{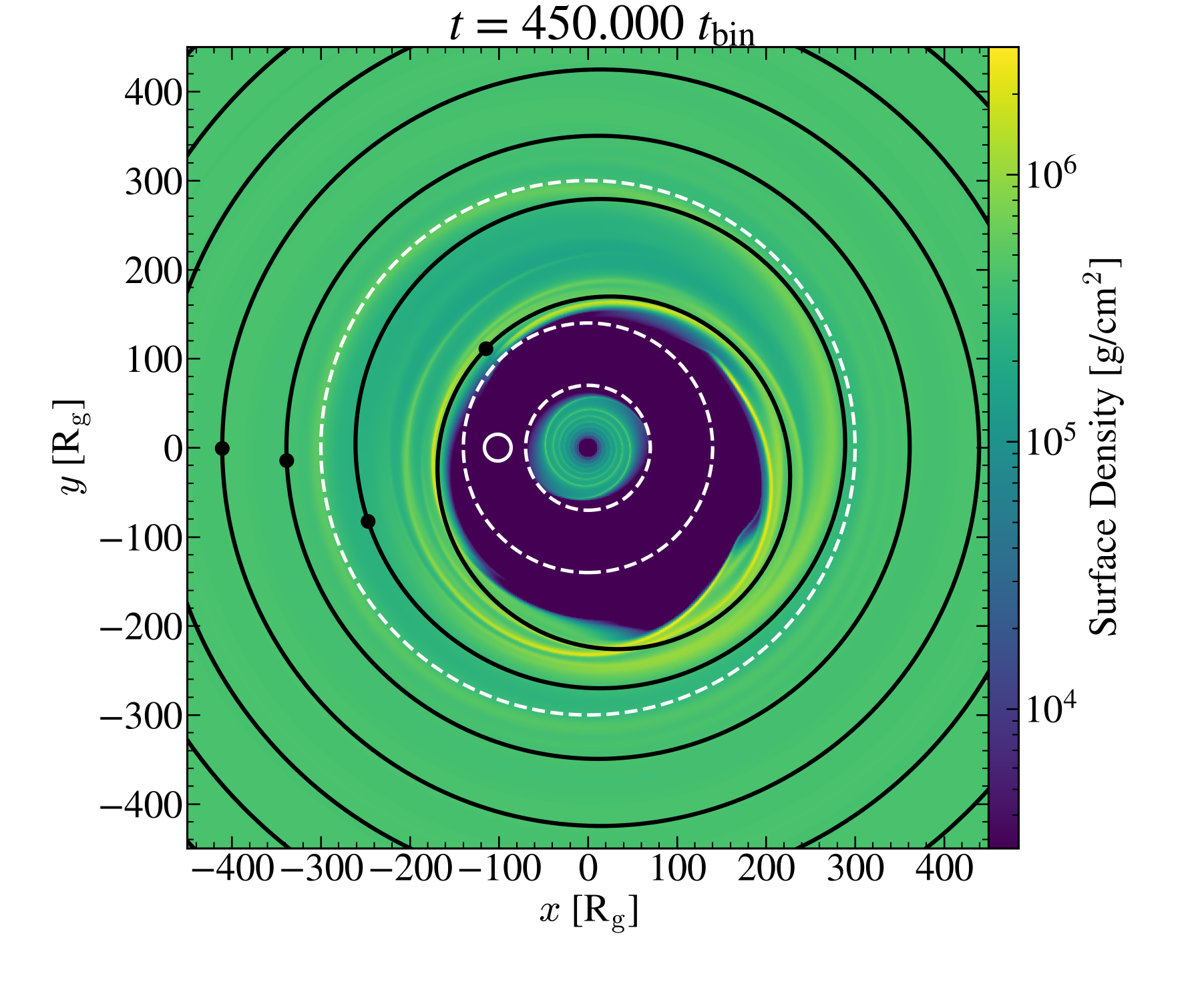}\hspace{-12pt}\includegraphics[trim= 2pt 24pt 2pt 2pt, clip,width=0.47\textwidth]{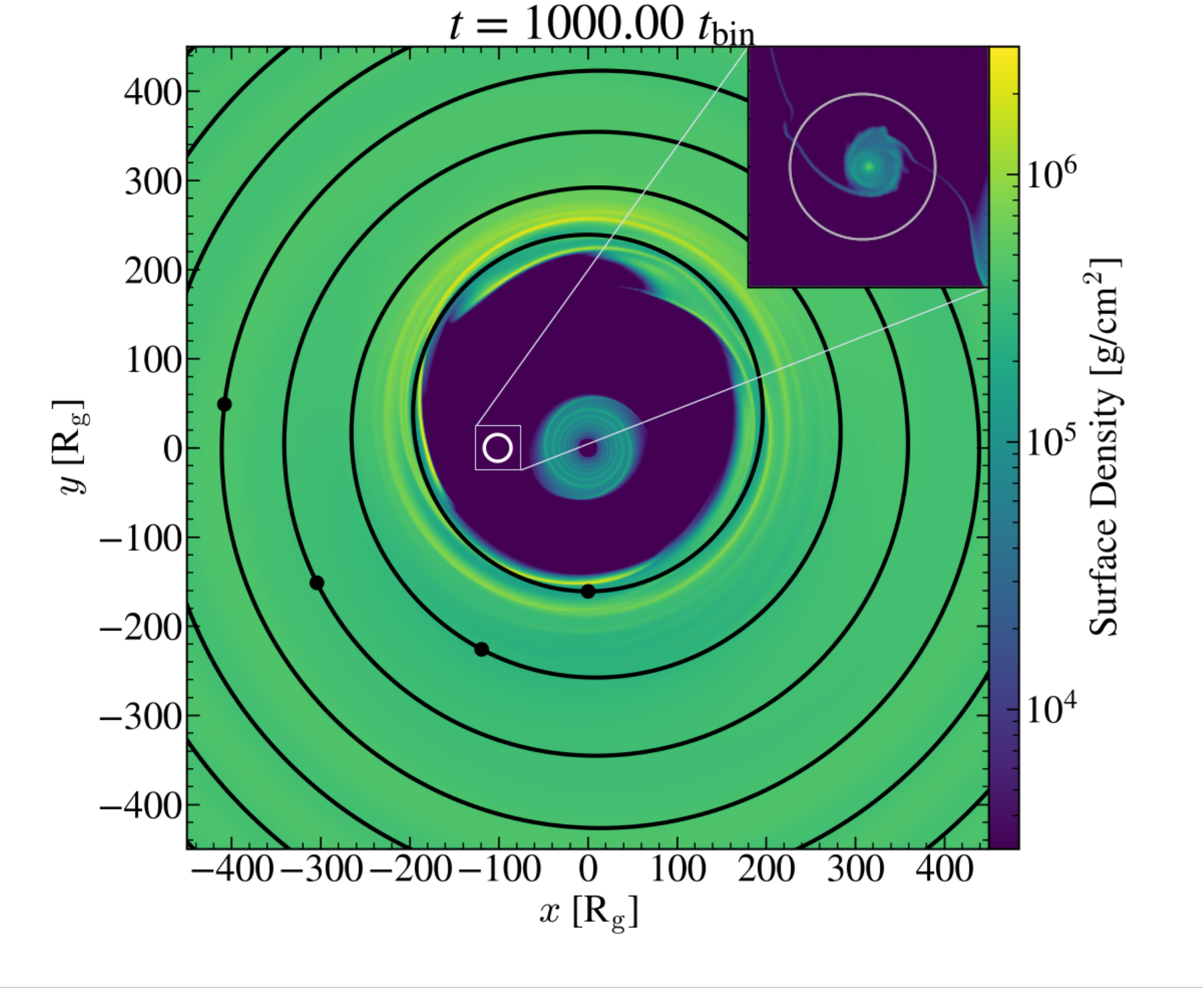}
    \caption{Snapshots of the gas surface density over time. The white circle indicates the extent of the secondary SMBH's Hill sphere, and the black ellipses show fits to the orbits taken by gas in the disk, with the circles on each ellipse indicating pericenter for that orbit. The snapshot at $t=1000\;t_{\rm{bin}}$ includes a $50\;R_g\times50\;R_g$ inset showing the minidisk and accretion streams, with the colorbar rescaled to span $3$--$3\times10^3$ g/cm$^2$ to better highlight the structure in this region. The snapshot at $t=450\;t_{\rm{bin}}$ has dashed lines showing the boundaries used in the torque analysis in \S\ref{sec:results:dynamics}. Early on, the gap opened by the secondary is circular, but interactions with the binary subsequently drive rapid eccentricity growth, which saturates between $400$--$500\;t_\mathrm{bin}$. The orientation of this eccentric gap precesses relative to the binary over long timescales.
    }
    \label{fig:dens_images}
\end{figure*}

In contrast to previous works at this mass ratio, we find that a $q=0.01$ binary is able to sustain a lopsided gap in the disk. In Fig.~\ref{fig:dens_images}, we show snapshots of the disk surface density at various times and overplot ellipses showing the orbits taken by gas in the disk. The ellipse-fitting procedure is based on that of \citet[][see their \S 4]{teyss17}. First, we calculate orbital elements for each cell in the simulation: the semi-major axis $a$, eccentricity $e$, and argument of pericenter $\overline{\omega}$. We then bin cells by semimajor axis and average the orbital properties of cells in the bin to obtain an overall fit to the orbit for a given value of $a$ and bin width $\delta a$. Throughout this analysis, we use a bin width of $\delta a = 1.0\; R_{\rm g}$.

\begin{figure}
    \center
    \includegraphics[width=0.47\textwidth]{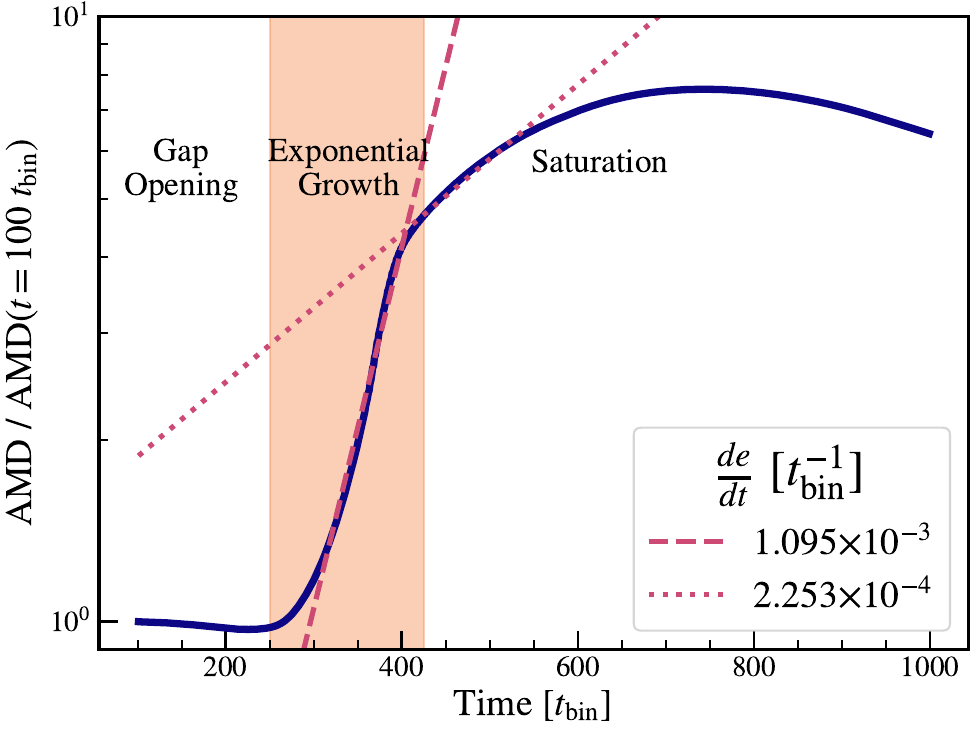}
    \caption{Total angular momentum deficit between $r=200\;R_\mathrm{g}$ and $r=400\;R_\mathrm{g}$. The orange box highlights the time during which the AMD grows exponentially. Linear fits to the growth rate are plotted for two separate times, one during the exponential phase and one immediately after as the AMD begins to saturate. These fits are labeled with the corresponding eccentricity growth rate, calculated by multiplying the slope of each fit by $\log(10)/(2\pi\times2)$, which corrects for the log scaling, the time unit, and the proportionality $\mathrm{AMD}\propto e^2$, as was done in \citet{teyss17}.} 
    \label{fig:amd}
\end{figure}

The eccentricity of the disk can be seen to grow over time and, as was done by \citet{teyss17}, we measure this growth rate using the disk's angular momentum deficit, 

\begin{equation}\label{eq:amd}
    \mathrm{AMD} = \int^{R_\mathrm{out}}_{R_\mathrm{in}}  \int_0^{2\pi} \Sigma R^2 \Omega \left(1 - \sqrt{1 - e^2}\right)R\;dR\;d\phi,
\end{equation}
 which is shown in Fig.~\ref{fig:amd}. Similar to \citet{kley06} and \citet{teyss17}, we observe an early relaxing phase where the gap is opened, followed by an exponential growth phase of the eccentric mode, and, finally, saturation. The saturation of the AMD indicates the settling of the large-scale dynamics of the system and coincides with the settling of behavior in other monitored quantities (see Section \ref{sec:em}), motivating our use of it as an indicator of reaching a quasi-steady state.

\begin{figure}
    \center
    \includegraphics[width=0.47\textwidth]{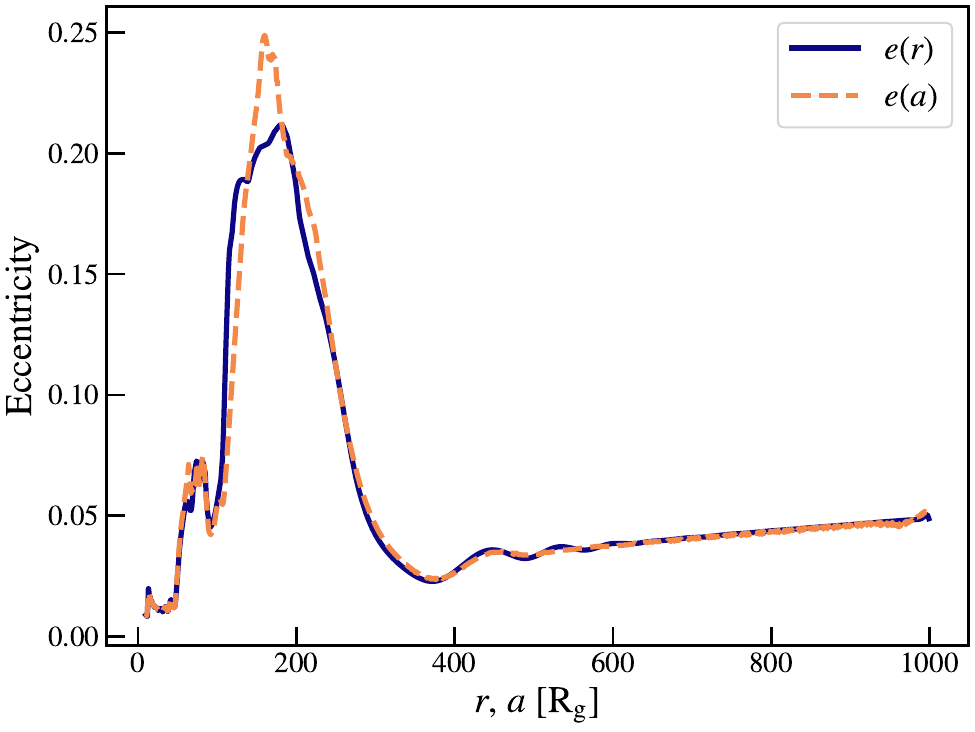}
    \caption{Eccentricity profile of the disk as a function of radial coordinate $r$ (solid purple) and semimajor axis $a$ (dashed orange). These profiles were time-averaged over $50\;t_\mathrm{bin}$ from $t=950\;t_\mathrm{bin}$ to $t=1000\;t_\mathrm{bin}$. The peak in eccentricity occurs at the approximate location of the gap edge.}
    \label{fig:ecc}
\end{figure}

\begin{figure}
    \center
    \includegraphics[width=0.47\textwidth]{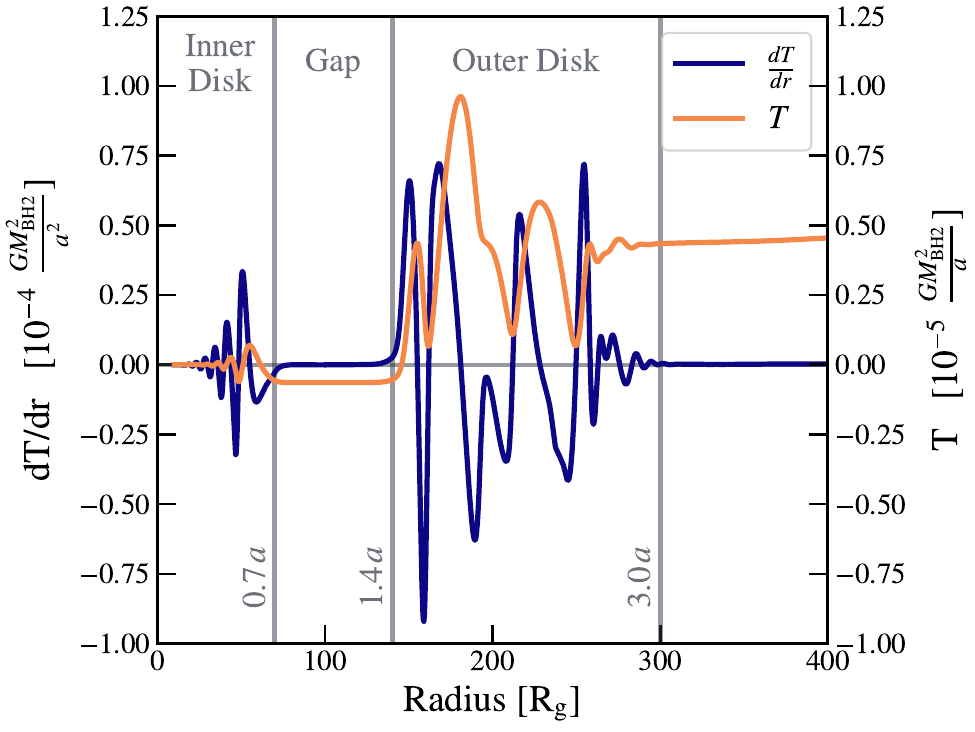}
    \caption{Torque density (purple) and integrated torque (orange) exerted by the binary on the disk. Quantities were time-averaged over 50 $t_\mathrm{bin}$ from $t=950\;t_\mathrm{bin}$ to $t=1000\;t_\mathrm{bin}$. The total torque on the disk is positive, and thus the reciprocal torque on the binary is negative, shrinking the binary separation over time. The disk can be separated into an inner, gap, and outer region based on the distribution of the torque density, indicated with the vertical grey lines at $0.7a$, $1.4a$, and $3a$. The total torque is plainly dominated by the contributions of the material near the outer disk edge.}
    \label{fig:torq}
\end{figure}

\begin{figure*}
    \center
    \includegraphics[width=0.94\textwidth]{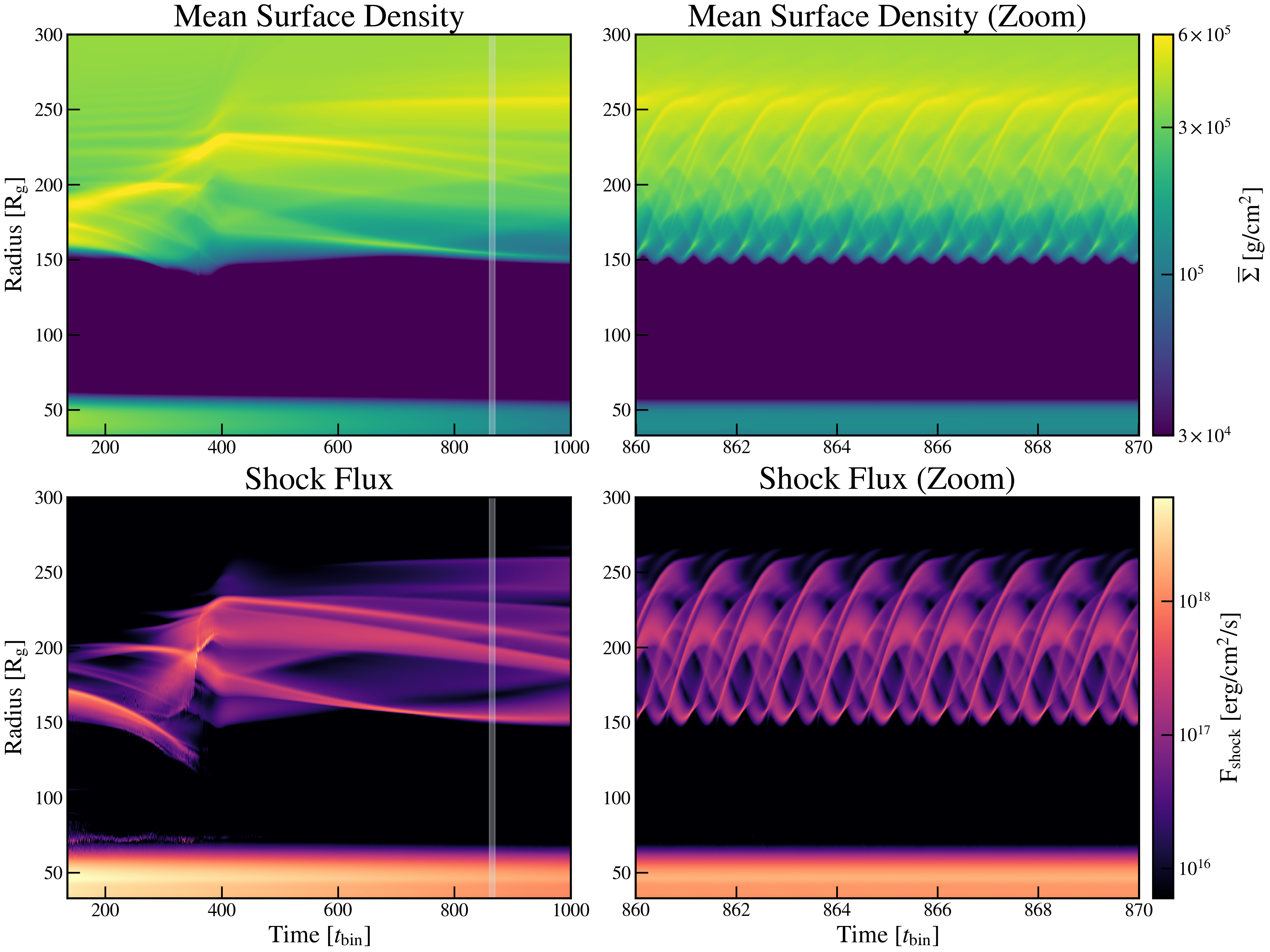}
    \caption{Azimuthally-averaged surface density $\bar{\Sigma}$ (\emph{top}) and azimuthally-summed shock flux $F_{\rm shock}$ (\emph{bottom}) as functions of time. The left column shows the quantities over the full simulation, while the right zooms in on 10 orbits during the quasi-steady state to highlight the parallel periodic structure in both quantities. The white bar on the left column plots indicates the times covered by the right column plots. The inner disk behavior is steady on short timescales, while a regular pattern is seen in both quantities at the outer edge of the gap ($\sim150$--$250\;R_{\rm g}$), indicating periodic shock waves excited by the binary. Shock dissipation in the inner disk gradually decreases over the course of the simulation, caused by depletion of surface density in the inner disk as gas is accreted across the inner boundary of the domain.
    }
    \label{fig:QvRvT}
\end{figure*}

\par Fig.~\ref{fig:ecc} shows the time-averaged eccentricity profile of our disk at the end of the simulation, both in terms of the radial coordinate of the simulation, $r$, and the semi-major axes, $a$, of the gas orbits. The gap edge, at approximately $r,a=200\;R_\mathrm{g}$, reaches an eccentricity of 0.25, larger even than the $e\approx0.1-0.2$ measured for higher mass ratios in past works with hotter disks \citep{farris14,mm08}. 

\par It can also be seen from Fig.~\ref{fig:dens_images} that the eccentric gap is precessing. Using the same orbit-fitting as was used to produce the orbital contours shown in Fig.~\ref{fig:dens_images}, we calculate the rotation angle of the $a=200\;R_{\rm g}$ ellipse, corresponding to the gap edge, over time. From this, we obtain a precession timescale of $\sim 1.41\times10^3\;t_\mathrm{bin}$, comparable to the $\sim 3.85\times10^3\;t_\mathrm{bin}$ calculated from the linear theory developed by \citet{teyss16}.

\par Fig.~\ref{fig:torq} shows the time-averaged torque density, defined in \citet{mm08} as
\begin{equation}
    \frac{dT}{dr} = -2 \pi R \left< \Sigma \frac{\partial \Phi}{\partial \phi}\right>,
\end{equation}
and integrated torque, $T(r)=\int^{r}_{0}\left({dT}/{dr}\right)dr$, of the secondary SMBH acting on the disk, averaged over 50 $t_\mathrm{bin}$. The strongest peak in the torque density occurs near $r=2.5a$, coinciding with the peak in the surface density distribution of the disk. Beyond this, the torque density quickly decays to oscillations about zero. The total torque experienced by the disk is positive, with $T(\infty)\approx7.89\times10^{50}\;{\mathrm{g\;cm^2}}{\mathrm{s^{-2}}}$. The reciprocal torque of the disk on the secondary is thus negative, serving to shrink the binary separation on a timescale $t_T = {M_2 a^2 \Omega_{\mathrm{BH2}}}/{T\left(\infty\right)}=3.62\times10^6\;t_\mathrm{bin}$. This is substantially longer than the runtime of the simulation, but much shorter than the timescales computed for the accretion of momentum and the emission of gravitational waves, implying that the binary's evolution is dominated by the gravitational torque of the disk at this time.

We can further explore where this negative torque comes from by following a similar procedure to \citet{tiede20} and \citet{munoz19}. We divide our domain into three regions based on the distribution of the torque density shown in Fig.~\ref{fig:torq}: (1) an inner region $R<0.7a$, consisting of the inner disk, (2) an outer region $R>1.4a$, which includes the entire outer disk, and (3) a gap region $0.7a<R<1.4a$. It can be seen from the integrated torque in Fig.~\ref{fig:torq} that the torque is dominated by the contributions from the outer disk, with $|T_{\mathrm{out}}|\approx9|T_{\mathrm{in}}|$, and, further, that these contributions becomes negligible past $R\sim3a$. We plot guides showing each of these regional boundaries in the bottom-left panel of Fig.~\ref{fig:dens_images}, which reveals that the outer region $1.4a<R<3a$ dominating the torque corresponds to the approximate extent of the overdense region near the eccentric gap edge. This is in line with the results of \citet{tiede20}, who found that the enhanced buildup of material near the gap edge in cold disks drives the total torque on the binary to be negative.

\begin{figure*}
    \center
    \includegraphics[width=0.94\textwidth]{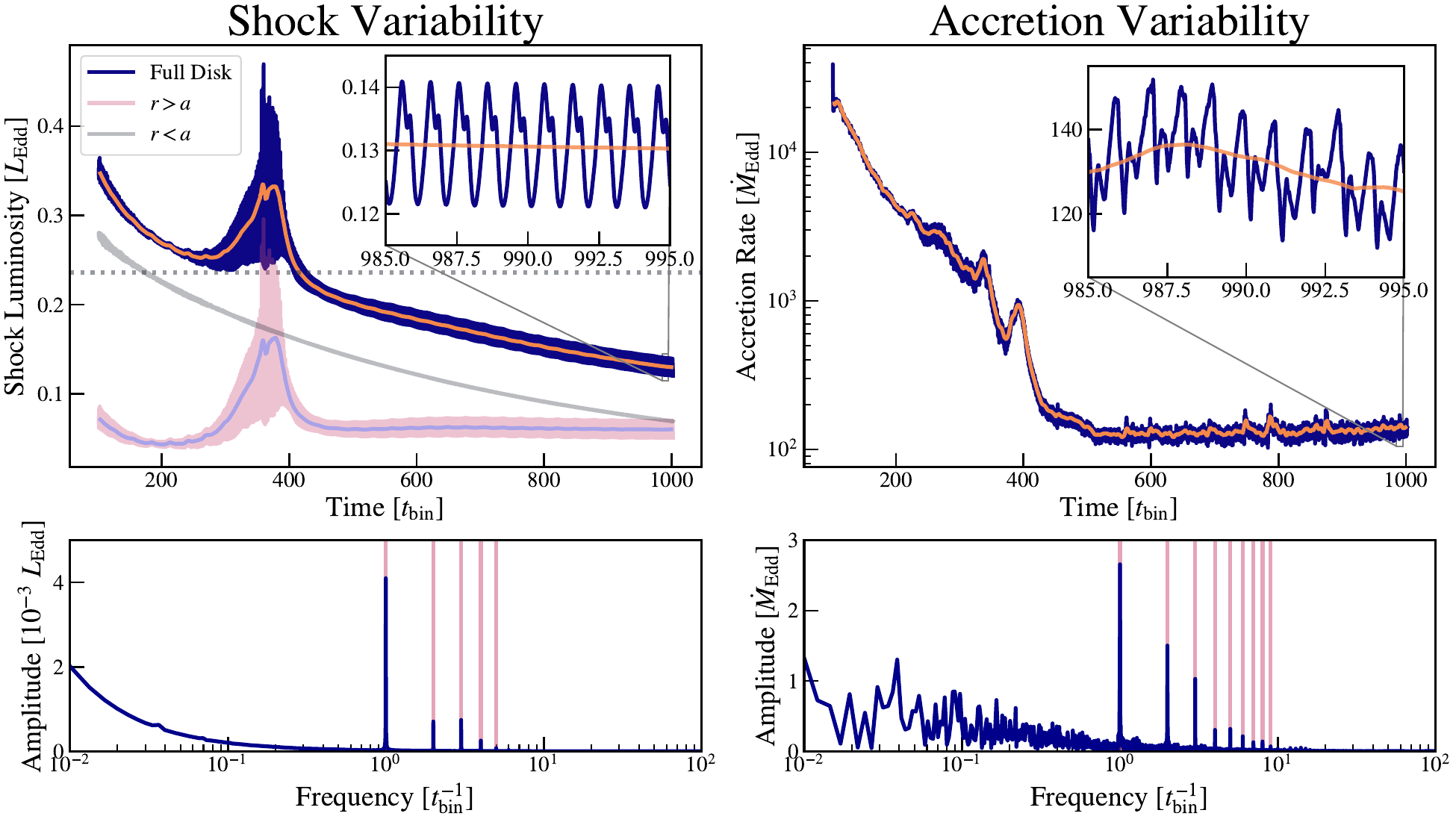}
    \caption{\emph{Top}: Shock dissipation rate summed over the disk (left) and accretion rate onto the secondary SMBH (right), shown over the full runtime of the simulation. The insets zoom in on 10 orbits late in the run to highlight the periodic structure of each quantity. The orange curves are moving averages of each quantity, taken over 10 orbits. For the shock lightcurve, the horizontal dotted line shows, for comparison, the thermal luminosity of the unperturbed disk model used for our initial conditions, though we note that the shocks will primarily emit in X-rays, while the thermal luminosity will peak at optical wavelengths. The pink and grey curves are, respectively, the shock luminosity from outside and inside the secondary's orbit, revealing that the variability is dominated by regions outside the secondary's orbit (see also Fig.~\ref{fig:QvRvT}) and that the behavior of shocks in that region have reached a steady state. The lilac curve is a moving-average of the $r>a$ shock luminosity, taken over 10 orbits. The large, long-duration flare in the shock luminosity between $300\;t_{\rm{bin}}$ and $400\;t_{\rm{bin}}$ coincides with the exponential growth in gap eccentricity illustrated in Fig.~\ref{fig:amd}. \emph{Bottom}: Fourier transforms of the shock lightcurve (left) and accretion rate (right). Only data from after the system reached a quasi-steady state around $\sim700\;t_{\rm{bin}}$ were included in these calculations. Both quantities exhibit clear periodicity on the orbital timescale of the binary.}
    \label{fig:varplots}
\end{figure*}

\subsection{Electromagnetic Emission}\label{sec:em}

The primary aim of this work is to investigate the time-varying electromagnetic properties of a low-mass-ratio SMBHB embedded in a realistic disk. To this end, we have employed two key proxies for variable electromagnetic emission generated from the system: shocks, tracked via dissipation in the artificial viscosity (\S\ref{sec:m:shocks}), and accretion rate onto the secondary (\S\ref{sec:m:accr}), standing in for the radiative emission of the minidisk. We find clear evidence for periodicity in both of these markers, each with a period matching the binary orbital period and a peak-to-trough ratio of $\sim1.2$.

\subsubsection{Shocks}

There are two major modes in which shocks occur in our disk: spiral shocks in the inner disk and shocks at the outer edge of the eccentric gap, driven by a portion of the gas streams feeding the secondary being flung back into the gap edge. The behavior of shocks in these regions can be observed in Fig.~\ref{fig:QvRvT}, which shows the surface density and shock dissipation flux as a functions of time and radius through the disk. We see that, in the quasi-steady state, the shock dissipation in the inner disk is steady, while the outer gap edge displays clear repeating shock waves, which map onto similar behavior in the surface density. It is these shock waves in the outer gap edge that drive variability in the total shock emission.

We produce a shock ``lightcurve" (Fig.~\ref{fig:varplots}, top left panel) by monitoring the total energy dissipated by artificial viscosity across the disk over our output timescale $\mathrm{DT}=10^{-3}\;t_\mathrm{bin}$. First, we note that the shocks are bright, producing an average luminosity $L_{\rm{shock}}\approx0.13\;\rm{L}_{\rm{Edd}}$ late in the simulation, with $\rm{L}_{\rm{Edd}}$ being the Eddington luminosity of the primary. For comparison, the unperturbed disk of our initial conditions has a luminosity $L_{\rm{disk}}\approx0.25\;\rm{L}_{\rm{Edd}}$, assuming an opacity dominated by electron scattering. 

The bottom-left panel of Fig.~\ref{fig:varplots} shows the Fourier transform of the lightcurve. There is a clear peak corresponding to the orbital frequency of the binary, with the first few harmonics of this frequency also present, but weak. This periodicity is plainly visible in the zoomed-in inset of the lightcurve itself, where we also note that the peak-to-trough ratio of the shock luminosity is only $1.17$, though this rises to $\sim1.4$ when considering emission from $r>a$ only, as may be relevant if the $r<a$ emission continues decaying to zero on longer timescales.

\subsubsection{Accretion}

The accretion rate onto the secondary, shown in the top-right panel of Fig.~\ref{fig:varplots}, is highly super-Eddington, hovering at around $\sim130\;\rm{\dot{M}}_{\rm{Edd}}$ in the steady state. Enhancement of accretion through the circumbinary disk and comparable accretion rates onto unequal-mass binary components has also been seen in preceding works such as \citet{farris14}. Enhancement of accretion through the disk is also, in general, expected given the enhancement of Reynolds stresses due to the secondary's perturbation of the flow.

The bottom-right panel of Fig.~\ref{fig:varplots} shows the Fourier transform of the secondary's accretion rate. Like the shock lightcurve, the accretion rate varies on the orbital frequency of the binary, with clear but weaker harmonics also apparent. The accretion rate is less steady from orbit to orbit than the shock lightcurve, but has a similar average peak-to-trough ratio of $\sim1.2$. The once-per-orbit peaks in accretion rate occur when the secondary makes its closest approach to the gap edge, stripping off material, some of which is accreted onto the minidisk, and the rest of which is flung back into the gap edge. Similar behavior is discussed in, e.g., \citet{dorazio13}.

We also monitored the linear and angular momentum accreted by the secondary to consider the resulting evolution of the secondary's orbit and spin, respectively. In the steady state, the secondary accreted momentum at a rate ${\Delta p}/{t_{\rm{bin}}} =1.01\times10^{-10}p_{\rm{BH2}}$, where $p_{\rm{BH2}}$ is the initial momentum of the secondary on its orbit. This net accretion torque is positive and therefore acts to increase the binary separation, but its magnitude is very low, validating our choice not to evolve the binary orbit based on the accreted momentum during the simulation. The accreted angular momentum is likewise inconsequential. Expressed as a rate of change of the dimensionless spin parameter, the secondary SMBH accretes angular momentum at a rate ${\Delta a}/{t_{\rm{bin}}} = 1.30\times10^{-10}$.

\section{Discussion}\label{sec:discussion}
In this section, we discuss the implications of our results in greater detail. First, we examine the disk dynamics and why the dynamics observed in this work differ from previous low-mass-ratio SMBHB simulations. Then, we discuss our proxies for electromagnetic variability and their implications for observational identification of SMBHB candidates. Finally, we look at the various torques acting on the binary and their implications for the binary orbital evolution.

\subsection{Disk Dynamics}\label{sec:disc:dynamics}
Previous SMBHB works have found that eccentric cavities as drivers of accretion variability are not present at low mass ratios, $q<0.05$ \citep{dorazio16}. Conversely, in this work, with the use of a self-consistent disk model for the surface density and scale height, we show that such behavior can be present in mass ratios as low as $q\sim0.01$. 

At a similar mass-ratio regime, simulations of super-Jupiters embedded in protoplanetary disks also produce eccentric gap systems with comparable eccentricity distributions to those in Fig. ~\ref{fig:ecc} \citep{dml21}.
They also find that the mass ratio for which eccentricity is excited coincides with the transition from inward to outward migration due to gravitational torques. Contrary to this, we are able to develop an eccentric gap while maintaining an inward migration of the secondary. Whether this decoupling of the two phenomena is due to our disk being much thinner than the thinnest $H/R$ tested in \citet{dml21}, our $H/R$ being non-uniform throughout our domain, or some other difference in the two setups is an interesting question, but beyond the scope of the present investigation.

The greatest difference between the disk in this study and those of preceding works is the disk model used for our initial conditions. In particular, our disk aspect ratio $H/R$ varies with radius and is, in general, much thinner than the constant $H/R=0.1$ disks common in the literature, with the disk having $H/R\approx10^{-2}$ at the location of the secondary (see Fig.~\ref{fig:ics}). The efficiency of gap-opening depends on the relationship between opposing torques: the tidal torque, $T_{\rm{tid}}\propto\left(H/R\right)^{-3}$, and the viscous torque $T_{\rm{visc}}\propto\nu\propto\left(H/R\right)^2$, which work to open the gap and to fill it in, respectively. In a colder, thinner disk, the tidal torque gets stronger and the viscous torque gets weaker, leading to wider, deeper gaps and allowing gap-opening to extend to SMBHBs with lower secondary masses \citep{crida06,duffell15,duffell20}. This phenomenon of wider gaps in colder disks has been observed in simulations of equal-mass binaries which test different values of $H/R$ \citep{ragusa16,tiede20}, and here we demonstrate that this behavior continues to lower mass ratios. 

So a colder, thinner disk makes gap-clearing more effective. In clearing a deeper gap, the secondary removes material near its orbit, weakening the damping of eccentricity occurring at eccentric corotation resonances (ECRs), enhancing the ability for the binary to drive eccentricity growth in the disk \citep{teyss17}. In summary, thinner disks are conducive to opening deeper gaps at lower mass ratios, which in turn drives eccentricity evolution in the outer disk and the consequent variable accretion onto the binary. Given that AGN have disks with $H/R\approx10^{-2}$--$10^{-3}$ \citep{krobook}, consistent with the disk model used in this study, this suggests that the electromagnetic variability signatures seen in this and previous works may exist for lower mass ratios than previously indicated, increasing the population of SMBHB candidates identifiable in observations.

\subsection{Variability}\label{sec:disc:variability}

The underlying motivation of this work and many other simulations of SMBHBs is to better characterize the periodic electromagnetic emission which can differentiate binaries from single-SMBH AGN. Previous works generally find that high-mass-ratio binaries exhibit periodic accretion rates, but that accretion becomes steady for $q\lesssim0.05$ \citep{farris14,dorazio16,duffel20}. The timescale of accretion variability depends on the mass ratio, with lower mass ratios $0.05\lesssim q\lesssim0.25$ being modulated only on the binary orbital timescale, while at larger mass ratios an overdense lump forms on the gap wall, causing periodic accretion on a timescale linked to its own orbital time \citep{farris14,dorazio13}.

In contrast, we find periodic accretion onto our secondary even with $q=0.01$, with a timescale matching the orbital timescale of the binary. This matches the properties seen for binaries in the $0.05\lesssim q\lesssim0.25$ regime in these past works, and implies that this regime may simply extend to lower mass ratios given a thinner disk. We note that the exact structure of this periodicity is dependent on the sink prescription used for the accretion. In the broadest sense, the accretion timescale used determines how quickly the minidisk around the secondary is drained. If this timescale is very short, the minidisk vanishes, and the amplitude of variability is magnified, being dependent only on the rate at which material enters the accretion zone. For slower sink accretion, where the minidisk persists over many orbits, the minidisk acts as a ``buffer," smoothing out the ``spiky" events of material being added to the minidisk (see \citet{duffel20} for one investigation of this relationship between accretion variability and sink accretion timescale). Here, we have chosen to model the minidisk as a steady $\alpha$-disk using the same model as for our initial conditions, which resulted in a stable minidisk with moderate accretion variability. Other disk models, such as an eccentric or shock-dominated disk, may be appropriate to consider for modeling the sink accretion timescale, but will likely impact the variability seen here primarily by being either faster or slower than the timescale used in this work, resulting in variability which is more or less pronounced, respectively.

Shocks have not been used as a signifier of variable EM emission in preceding isothermal works, though they have been shown to be a source of periodic X-rays in non-isothermal simulations such as \citet{tang18}, implying that shock capturing and characterization is necessary to understand a potentially crucial source of periodic emission driven by the binary. We find from our shock monitoring scheme that, like \citet{tang18}, shocks occur periodically from rejected accretion streams striking the gap wall. This periodic shocking, like our variable accretion, occurs on a timescale equal to the orbital timescale of the binary.

\begin{figure*}
    \center
    \includegraphics[width=0.94\textwidth]{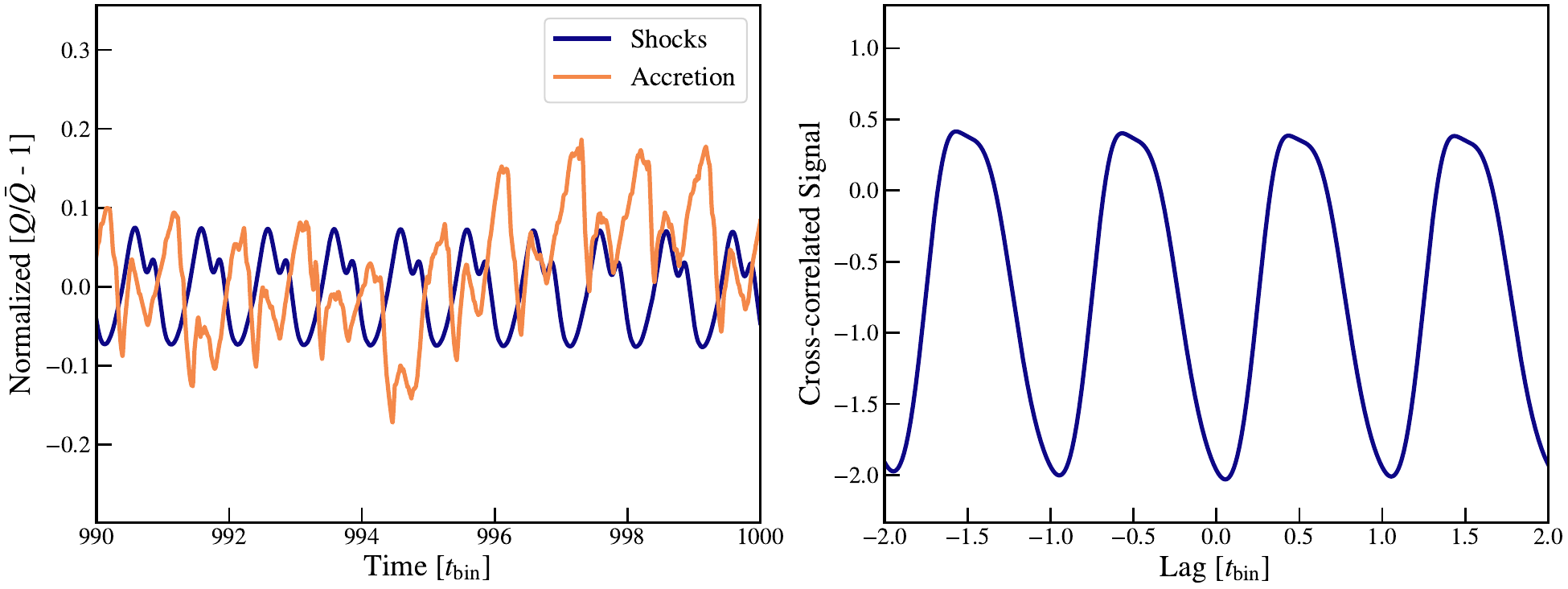}
    \caption{Examination of the timing relationship between shocks and accretion. \emph{Left:} Shock lightcurve (\emph{purple}) and accretion rate (\emph{orange}), each normalized by their mean value. \emph{Right:} Cross-correlation of the shock and accretion ``lightcurves." The cross-correlation is itself periodic due to both signals being periodic, and peaks at a lag of $0.43\;t_\mathrm{bin}$. Since the two signals have a period of $t_\mathrm{bin}$, this affirms that they are almost fully out of phase with one another, as is apparent by eye in the left panel.}
    \label{fig:lc_v_mdot}
\end{figure*}

One important caveat to the variability seen in our accretion and shocks is that the amplitude is modest, with both proxies peaking $\lesssim10\%$ above their mean. However, the timescale is well-constrained and short ($t_{\rm{bin}}\approx0.1\;\rm{yr}$), allowing many cycles of such a system to be obtained from continued monitoring with UV or X-ray instruments, alleviating the difficulties discussed by \citet{vaughan16} in distinguishing genuine periodicity from insufficiently-sampled red noise.

Further still, in Fig.~\ref{fig:lc_v_mdot} we examine the timing relationship between the shock lightcurve and the secondary's accretion rate via cross-correlation and find that they are nearly fully out-of-phase with one another, with a timing offset of $\Delta t_{\rm{lag}}\approx0.43\;t_{\rm{bin}}$ between peaks in the accretion rate and peaks in the shock luminosity. This parallels the relationship found between the optical lightcurve and accretion rate for a circular binary in the fully adiabatic simulations performed by \citet{ws22}. This timing offset is potentially valuable as an identifier of SMBHB candidates, as the shocks can produce X-ray variability \citep{tang18}, while the minidisk emission will typically peak at UV or longer wavelengths, depending on the mass of the secondary \citep{ss73}. This type of correlated variability across wavebands is not expected for red noise, and thus cross-correlation analysis between UV and X-ray monitoring of AGN could be a valuable observational pursuit for identifying candidate SMBHBs.

\subsection{Binary Orbital Evolution}\label{sec:disc:binorb}

A recurring point of discussion around SMBHBs is the sign of the gravitational torque acting on the binary, as this tends to dominate the evolution of the separation prior to the gravitational wave-dominated phase. We find that our disk exerts a negative torque on the binary, reducing the binary separation over time, which matches \citet{duffel20}, which finds that the torque is negative for mass ratios below $q\sim0.05$, albeit for a much warmer disk with constant $H/R=0.1$. The majority of the negative torque can be attributed to the build up of material at the outer edge of an eccentric gap -- as seen in the case of equal-mass binaries in \citet{tiede20} in which this effect is greater in magnitude for colder disks that accumulate more material at the gap edge. Conversely, \citet{derd21} find, for lower mass ratios than ours, that the torque on the binary flips from negative to positive as the disk becomes thinner, though they note that the density asymmetry key to determining the gravitational torque is unresolved, falling within the smoothing length of the secondary in some of their simulations. In general, these works and ours suggest that the magnitude of this effect is dependent on the specific disk model as well as the mass ratio of the embedded binary, likely due to whether the system leads to the opening of an eccentric gap. A systematic study of the gravitational torque over a wide range of gap opening scenarios is an interesting avenue for future investigation.

While not explored in this work, it is expected that, just as the binary excites eccentricity in the disk, the disk should excite eccentricity in the binary orbit. There are several works which have explored the case of eccentric binaries \citep{dorazio21,miranda17} finding that the eccentricity of the binary affects the variability of accretion, the morphology and dynamics of the disk gap, and the evolution of the binary orbit. \citet{franchini23} have also found that fixing the binary orbit leads to overestimation of the gravitational torque and underestimation of accretion torques for equal-mass binaries. Exploring the effects of eccentric and live binaries in our cold disk model and how these compare to the existing warmer disk works is an interesting path for future investigation.

\section{Conclusions}\label{sec:conclusion}

We performed a 2D, locally isothermal hydrodynamic simulation of a low-mass-ratio ($q=10^{-2}$) SMBHB embedded in a disk with initial surface density and sound speed profiles derived from a physically self-consistent disk model. We monitored the dynamical evolution of the system as well as two proxies for electromagnetic variability, the accretion rate onto the secondary SMBH, and the energy dissipated by shocks. 

Our main findings can be summarized as follows.
\begin{enumerate}
    \item The binary opens a wide, eccentric gap which precesses on long timescales. This behavior is seen throughout the literature at larger mass ratios, but has not generally been observed for $q\lesssim0.05$. We attribute this change to our disk model, which is much thinner than the $H/R=0.1$ disks common in the existing literature, and in line with the $H/R\sim0.01$--$0.001$ expected for real AGN disks. A thinner disk is both more susceptible to gravitational torques from the binary, which open the gap, and experiences weaker viscous torques, which serve to close it. Wider cavities in thinner disks have been seen in a few works which explored changing $H/R$ with $q=1$ binaries, and here we have shown that this behavior extends to $q=0.01$ binaries. 
    \item We find that accretion onto the secondary SMBH is clearly variable, with a period matching the orbital period of the binary and a peak-to-trough ratio of $\sim1.2$. As with the eccentric gap, this behavior was previously not seen for $q\lesssim0.05$ in works using thicker disks. We find, as has been the case in previous works, that the peaks in the accretion rate occur due to the passage of the secondary near the overdensity at the edge of the gap, stripping off material to feed the minidisk. This process is necessarily linked to the gap becoming eccentric. Since we find that a thinner disk leads to eccentricity at smaller mass ratios than previously seen, it is unsurprising that accretion variability follows. Importantly, since real AGN disks are expected to be thin ($H/R\sim0.01$--$0.001$), we expect that the parameter space for which real binaries are variable is wider than has been previously established by works with thicker disks.
    \item We find that shocks excited by the binary are also periodic, with a period matching the orbital period of the binary and a peak-to-trough ratio of $1.17$. Shocks have been found to be an important source of periodic X-ray emission in non-isothermal works, but have not previously been monitored in isothermal simulations.
    \item We find that there is a correlated lag between the accretion and shock lightcurves. The two quantities are nearly fully out of phase, with the shocks lagging behind accretion by $0.43\;t_\mathrm{bin}$. This presents a potential smoking gun for binary candidacy in observations. Since accretion tracks the minidisk luminosity, which will typically be brightest at ultraviolet wavelengths, and shocks are seen to be bright in X-rays, the timing correlation between shock dissipation and accretion rate implies correlated variability in separate wavebands. Observations can then be made to search for such correlated variability as a sign of a possible SMBHB rather than more ambiguous single-waveband variability.
\end{enumerate}

\section*{Acknowledgements}

This research was supported in part through computational resources and services provided by Advanced Research Computing at the University of Michigan, Ann Arbor. This work was funded in part by Michigan Space Grant Consortium, NASA grant \#NNX15AJ20H.

A.K. would like to acknowledge support provided by NASA through the NASA Hubble Fellowship grant \#HST-HF2-51463.001-A awarded by the Space Telescope Science Institute, which is operated by the Association of Universities for Research in Astronomy, Incorporated, under NASA contract NAS5-26555. 

M.R. acknowledges support from NASA grants 80NSSC20K1541 and 80NSSC20K1583 and NSF grants AST-1715140 and AST-2009227.

\section*{Data Availability}

The data underlying this article will be shared on reasonable request to the corresponding author



\bibliographystyle{mnras}
\bibliography{references} 




\appendix

\section{Disk Model}\label{appendix:disk}
\renewcommand{\thefigure}{A\arabic{figure}}
\setcounter{figure}{0}

\renewcommand{\theequation}{A\arabic{equation}}
\setcounter{equation}{0}

The motivation behind the disk model used in this work is to have a model which is both consistent with the equations describing a steady disk as well as intuitively tunable in regards to where gas and radiation pressure dominate in the disk. For this, we start, as indicated, from the equations for a steady disk, as presented in \citet{fkr}:

\begin{equation}\label{app:steadydisk}
    \left.
    \begin{aligned}
        &\mathrm{a}.\;\;\rho = \Sigma / H;\\
        &\mathrm{b}.\;\;H = c_s \Omega_K;\\
        &\mathrm{c}.\;\;c_s^2 = P/\rho;\\
        &\mathrm{d}.\;\;P = \frac{\rho k_B T_c}{\mu m_p} + \frac{4\sigma}{3c} T_c^4;\\
        &\mathrm{e}.\;\;\frac{4\sigma T_c^4}{3\tau} = \frac{3GM\dot{M}}{8\pi R^3}\left[1 - \left(\frac{2R_g}{R}\right)^{1/2}\right];\\
        &\mathrm{f}.\;\;\tau = \Sigma \kappa_{\rm R}\left(\rho,T_c\right)=\tau\left(\Sigma,\rho,T_c\right);\\
        &\mathrm{g}.\;\;\nu\Sigma = \frac{\dot{M}}{3\pi} \left[1-\left(\frac{2R_g}{R}\right)^{1/2}\right];\\
        &\mathrm{h}.\;\;\nu = \nu\left(\rho,T_c,\Sigma,\alpha,...\right),
    \end{aligned}
    \right\},
\end{equation}

The symbols used throughout these equations are all defined in \S \ref{methods:setup} of the main text.

For convenience, we introduce the convention $f\equiv[1-\left(2R_g/R\right)^{1/2}]$. We also take that our disk's kinematic 
viscosity is represented by a conventional $\alpha$-viscosity, of the form
\begin{equation}\label{app:visc}
    \nu = \alpha c_s H.
\end{equation}

Finally, we introduce a parameterization of the scale height $H$, separating the contributions from gas and radiation pressures:
\begin{equation}\label{app:Hdef}
    H^2 \equiv H_g^2 + H_r^2,
\end{equation}
where
\begin{equation}
\begin{aligned}
    &H_g^2 \equiv \frac{P_g}{\rho \Omega_K^2},\\
    &H_r^2 \equiv \frac{P_r}{\rho \Omega_K^2}.
\end{aligned}
\end{equation}
and 
\begin{equation}
\begin{aligned}
    &P_g = \frac{\rho k T_c}{\mu m_p},\\
    &P_r = \frac{4\sigma}{3c}T_c^4.\\
    \end{aligned}
\end{equation}

Then, for $H_r$, we have
\begin{equation}
    H_r^2 = \frac{4\sigma}{3c}T_c^4 \rho^{-1} \Omega_K^{-2}.
\end{equation}
Then, combining this with equations~(\ref{app:steadydisk}e), (\ref{app:steadydisk}f), and (\ref{app:steadydisk}a), we obtain
\begin{equation}\label{app:hrbeta}
\begin{aligned}
    H_r^2 &= \frac{3}{8\pi c} \kappa_{\rm R} \dot{M} H f\\
    &= h H,\\
\end{aligned}
\end{equation}
where, for convenience, we have defined 
\begin{equation}\label{app:beta}
    h\equiv 3\kappa_{\rm R} \dot{M} f / (8\pi c).
\end{equation}
We note here that far from the central object, where $R\gg 2R_g$, $f\approx 1$ and so, for $\kappa_{\rm R}=\rm{constant}$, as is chosen for this work, $h$ also goes to a constant. Further, when radiation pressure dominates, $H\approx H_r \Rightarrow H_r\approx h$. This combination of facts allows $h$ to be interpreted as a kind of ``minimum scale height" of the disk, whose value is set by the accretion rate $\dot{M}$.

Squaring both sides of equation~(\ref{app:hrbeta}) and recalling our definition of $H$ in equation~(\ref{app:Hdef}),
\begin{equation}
    H_r^4 = h^2\left(H_g^2+H_r^2\right).
\end{equation}
The only positive-valued solution for $H_r$ from this is
\begin{equation}
    H_r^2 = \frac{h^2}{2}\left(1 + \sqrt{1 + 4h^{-2}H_g^2}\right).
\end{equation}

Following convention, we choose to parameterize $H_g$ as a powerlaw in $R$,
\begin{equation}
    H_g\equiv h \left(\frac{R}{R_c}\right)^k.
\end{equation}
Since $h\approx\rm{constant}$ at large R, which is where gas pressure dominates, this expression effectively prescribes $H_g$ as a powerlaw of slope $k$. $R_c$ sets the radius at which $H_g=h$. Since $H_r\approx h$ in the radiation-dominated region, $R_c$ then determines the approximate radius where the disk transitions from radiation-dominated to gas-dominated.

Taking this expression for $H_g$, $H_r$ now becomes
\begin{equation}
    H_r^2 = \frac{h^2}{2}\left(1 + \sqrt{1 + 4\left(\frac{R}{R_c}\right)^{2k}}\right),
\end{equation}
and for the total scaleheight $H$ we obtain
\begin{equation}\label{app:Hfinal}
    H^2 = h^2 d\left(\frac{R}{R_c}\right),
\end{equation}
where we define
\begin{equation}
    d\left(\frac{R}{R_c}\right) \equiv \left[\left(\frac{R}{R_c}\right)^{2k} + \frac{1}{2}\left(1 + \sqrt{1 + 4\left(\frac{R}{R_c}\right)^{2k}}\right)\right]
\end{equation}

From here, we are ready to solve for the properties of our disk. We start with surface density by combining equation~(\ref{app:visc}) with equations~(\ref{app:steadydisk}b) and (\ref{app:steadydisk}g) to get
\begin{equation}\label{app:dens1}
    \Sigma = \frac{1}{3\pi}\alpha^{-1}\Omega_K^{-1} H^{-2} \dot{M} f\\.
\end{equation}

We can likewise obtain an expression for the midplane temperature by combining equations~(\ref{app:steadydisk}e) and (\ref{app:steadydisk}g),
\begin{equation}\label{app:temp1}
    T_c^4 = \frac{9G}{32\pi\sigma}M\dot{M}R^{-3}\kappa{\rm R} \Sigma f.
\end{equation}
Then, substituting in our expression for $\Sigma$ from equation~(\ref{app:dens1}),
\begin{equation}\label{app:temp2}
    T_c^4 = \frac{3G}{32\pi^2\sigma}\alpha^{-1}\kappa_{\rm R}M\dot{M} \Omega_K^{-1} H^{-2} f^2.
\end{equation}

We obtain the sound speed from combining equation~(\ref{app:Hfinal}) with equation~(\ref{app:steadydisk}b),
\begin{equation}\label{app:cs1}
    c_s = \frac{3}{8\pi c}\kappa_{\rm R} \dot{M} \Omega_K d\left(\frac{R}{R_c}\right) f.
\end{equation}

Finally, we choose, for convention, to recast equations~(\ref{app:beta}), (\ref{app:dens1}), (\ref{app:temp2}), and (\ref{app:cs1}) in terms of $r\equiv R/R_g$, $m\equiv M/M_\odot$, and $\dot{m}\equiv\dot{M}/\dot{M}_{\rm{Edd}}$. $R_g\equiv GM/c^2$ is the gravitational radius of the central object, $M_\odot$ is the mass of the Sun, and $\dot{M}_{\rm{Edd}}\equiv 4\pi GMm_p/(\eta \sigma_T c)$ is the Eddington accretion rate of a pure hydrogen gas onto the central object with an accretion efficiency $\eta$. Taking these definitions and also that $\Omega_K\equiv\sqrt{GM/R^3}$, we obtain
\begin{equation}\label{app:minheight}
    h \equiv \frac{3}{2} \frac{GM_\odot}{\eta c^2} m\dot{m} f,
\end{equation}
\begin{equation}\label{app:dens}
    \Sigma(r) = \frac{16}{27}\frac{m_p}{\sigma_T} \eta \alpha^{-1} \dot{m}^{-1} r^{\frac{3}{2}} \left(1 - \sqrt{\frac{2}{r}}\right)^{-1} d\left(\frac{r}{r_c}\right)^{-1},\\
\end{equation}
\begin{equation}
    T_c^4(r) = \frac{2c^5}{3\sigma G M_\odot}\kappa_{\rm R}^{-1} \alpha^{-1} m^{-1} r^{-3/2} d\left(\frac{r}{r_c}\right)^{-1},\\
\end{equation}
\begin{equation}\label{app:soundspeed}
    c_s(r) = \frac{3}{2}c\eta^{-1}\dot{m}r^{-3/2}\left(1 - \sqrt{\frac{2}{r}}\right)^2 d\left(\frac{r}{r_c}\right)^{1/2},\\
\end{equation}
where $r_c\equiv R_c/R_g$. Equations~(\ref{app:minheight}), (\ref{app:dens}), and (\ref{app:soundspeed}) are then, respectively, the same as equations~(\ref{eq:minheight}), (\ref{eq:density}), and (\ref{eq:soundspeed}) presented in the main body of this work.


\bsp	
\label{lastpage}
\end{document}